\documentclass[twocolumn]{rps-esrel2021}

\def\papername{\jobname}

\usepackage[authoryear]{natbib}
\usepackage{amsmath}
\usepackage{amssymb}
\usepackage{xcolor}

\begin{document}

\markboth{Hamad, Han, and U\c{c}an}{Online Estimation of Overload Risk in 5G Multi-Tenancy Network}

\twocolumn[

\title{Online Estimation of Resource Overload Risk in 5G Multi-Tenancy Network}

\author{Yasameen Shihab Hamad}

\address{Faculty of Engineering and Natural Sciences, Altinbas University, Turkey. \email{yasameen.hamad@ogr.altinbas.edu.tr}}

\author{Bin Han}

\address{Division of Wireless Communications and Radio Positioning, University of Kaiserslautern, Germany. \email{binhan@eit.uni-kl.de}}

\author{Osman Nuri U\c{c}an}

\address{Faculty of Engineering and Natural Sciences, Altinbas University, Turkey. \email{osman.ucan@altinbas.edu.tr}}

\begin{abstract} 
The technology of network slicing, as the most characteristic feature of the fifth generation (5G) wireless networks, manages the resources and network functions in heterogeneous and logically isolated slices on the top of a shared physical infrastructure, where every slice can be independently customized to fulfill the specific requirements of its devoted service type. It enables a new paradigm of multi-tenancy networking, where the network slices can be leased by the mobile network operator (MNO) to tenants in form of public cloud computing service, known as Slice-as-a-Service (SlaaS). Similar to classical cloud computing scenarios, SlaaS benefits from overbooking its resources to numerous tenants, taking advantage of the resource elasticity and diversity, at a price of risking overloading network resources and violating the service-level agreements (SLAs), which stipulate the quality of service (QoS) that shall be guaranteed to the network slices. Thus, it becomes a critical challenge to the MNOs, accurately estimating the resource overload risk - especially under the sophisticated network dynamics - for monitoring and enhancing the reliability of SlaaS business. 

While existing literature are mostly considering over-simplified models of slice resource load, in this paper we propose a novel approach assuming a Markov-Truncated-Gaussian-mixture model, which is capable of accurately approximating any practical resource load distribution. The proposed approach is demonstrated by numerical simulations to be effective and accurate.
\end{abstract}


]

\section{Introduction}\label{sec:intro}
Network slicing, as summarized by \cite{song2019hierarchical}, is considered as a characterizing feature of 5G networks, which proposes to divide a physical network into customizable slices, thus allowing to fulfill different requirements of heterogeneous network service scenarios. It has been built up as an efficient solution that reduces the cost of supporting a wide range of different services and applications. Under the network slicing concept, as concluded by \cite{kaloxylos2018survey} and \cite{ai2020joint}, the physical network is divided into numerous virtual networks, known as the network slices, each consisting of set resources and network functions. Every slice is constructed as an end-to-end virtualized network and is designed in specific terms of resources to achieve different requirements of the services, as described by the \cite{alliance2016description}.
 
Network slicing allows to provide telecommunication networking as a public cloud service, where network slices can be created upon service demand and rented to different tenants. The motivation of establishing a multi-tenancy system in 5G communication is that to enable a shared economy, where different tenants flexibly share the same network infrastructure, which significantly reduces both the capital expenditure (CAPEX) and the operations expenditure (OPEX). But on the other hand, as pointed out by \cite{bezemer2010enabling}, new challenges are also exacerbated in the meanwhile. Generally, the slices rented to different tenants must be isolated from each other for independent operations and independent performances. Such isolation requires the establishment of new virtual machines and logical networks for every individual tenant, and thus leads to a more complex network topology and new technical issues in the networking system, as foreseen by \cite{li2020iot}. 

Among all challenges raised by multi-tenancy network slicing, resource elasticity is one of the most critical. As shown by \cite{zanzi2018ovnes} and \cite{salvat2018overbooking}, the principle of shared economy tend to encourage the mobile network operators (MNOs) to overbook their resources to more tenants, so that a higher resource utilization rate and therewith a higher profit can be achieved by means of exploiting the inter-tenant load diversity. However, as pointed out by \cite{han2019utility}, it also takes the cost of increased risk of resource overload, which will lead to inevitable violations of the service-level agreements (SLAs), which generate penalties and losses of reputation in the market. Thus, the risk overload has to be accurately assessed to help select an appropriate policy of network slice admission control. Moreover, as discussed by \cite{barona2017towards}, due to the dynamic nature and different requirements of users, the resource load of a slice highly depends on the user behavior patterns of connection, data traffic, and mobility. Since such patterns commonly vary over time with a significant randomness, we need an agile and smart mechanism to monitor and predict the network resource load in an online fashion.

In this paper, we propose a mechanism to predict the resource overloading risk in sliced 5G networks. More specifically, we consider a Markov-Truncated-Gaussian-mixture model of the network resource load, which consists of a Markov chain to describe the policy-based admission control mechanism under arbitrary consistent policy, and a (truncated) Gaussian-mixture model that is capable of approximating any resource load distribution of a slice with satisfactory accuracy. Based on this model, we design a multi-stage overload risk estimation procedure. Our approach begins with a online feature extraction that roughly estimates the birth/death rates of network slices and long-term load distribution of every slice. These extracted features are then fed into two parametric estimators for the Markov model of active slice number, and the Gaussian-mixture model of single-slice-load, respectively. Both estimated models are then combined to compose a high-ordered Gaussian-mixture model for the resource load, from which the outage risk is obtained.

The remainder of this paper is organized as follows. We start with 
the problem model in Section \ref{sec:setup}, based on which we analyze the overload risk model in Section \ref{sec:analysis}, and propose an estimation method in Sec. \ref{sec:approach}. In Section \ref{sec:simulation} we present our effort and the results of numerical simulations for evaluating our proposals. Finally, conclusions and future work are presented in Section \ref{sec:conclusion}.

\section{Problem Setup}\label{sec:setup}
We consider a sliced multi-tenancy 5G system, where new slices are created upon randomly arriving tenant requests and the MNO's admission control policy. More specifically, requests for creating new slices are randomly generated by numerous tenants, arriving as a Poisson process. These requests are buffered in a First-Come-First-Served (FCFS) queue, and an admission control module makes the decision if to admit the first request in queue and create a new slice, or to keep it awaiting in the buffer. Furthermore, existing slices are also randomly released upon Poisson service termination. This service scenario and admission control mechanism has been deeply investigated in our previous studies (see \cite{han2018markov, han2019utility, han2020multiservice}). Since the optimization of multi-service admission control policy is out of our focus in this paper, for simplification we consider here only one slice type. Especially, according to \cite{han2018markov}, given any consistent admission control policy, the admission event will also be Poisson. In our discussion below, we neglect the admission control module, simply consider a Possion distributed ``effective'' arrival process of the admitted requests, and let $\lambda$ denote the arrival rate. Also we consider the lifetime of all slices as geometrically distributed, with the average value $1/\eta$.

Thus, at any time, the set of active slices $\mathcal{S}$ becomes a random variable which is jointly determined by the stochastic processes of birth (admission) and death (release), which is a well-studied classical queuing model. Regarding the practical deployment, we consider a maximal allowed number of active slices $n_{\text{max}}$. Furthermore, as we have discussed in Section \ref{sec:intro}, the instantaneous resource load of every slice $s\in\mathcal{S}$ is also a random variable. Here we consider the resource loads of all slices to be i.i.d. according to an \emph{unknown} Gaussian-mixture distribution left-truncated at $0$ (since the resource load cannot be negative). Thus, the resource overload risk can be represented as 
\begin{equation}\label{eq:overload_risk}
	P_{\text{o}}=\text{Prob}\left(r_\Sigma>r_{\text{max}}\right),
\end{equation}
where $r_\Sigma\triangleq\sum\limits_{s\in\mathcal{S}}r_s$ and $r_\text{max}$ is a pre-defined threshold.

\section{Analysis}\label{sec:analysis}
We start from a meticulous investigation on the probability distribution of $r_\Sigma$. Since the birth-death process $\mathcal{S}$ is independent from the resource load of every existing slice $r_s$, clearly $f_{r_\Sigma}(x)$ is composed of a linear set of probability density functions upon different status $\mathcal{S}$:
\begin{equation}
	f_{r_\Sigma}(x)=\sum\limits_{n=0}^{n_{\text{max}}}f_{n}(x)P_n,
\end{equation}
where $P_n=P(\vert\mathcal{S}\vert=n)$ is the probability that $n$ slices are active, and
\begin{equation}
	f_n(x)\triangleq \left.f_{r_\Sigma}(x~\right\vert~\vert\mathcal{S}\vert=n).
\end{equation}
Since $r_s$ is an i.i.d. Gaussian mixture random variable left-truncated at $r_s=0$ for all $s\in\mathcal{S}$, it is easy to derive that $r_\Sigma$ is also a Gaussian mixture random variable left-truncated at $r_s=0$ for all $\vert\mathcal{S}\vert\in\mathbb{N}^+$, which we omit here.

Obviously, with all parameters consistent, $r_\Sigma$ is guaranteed stationary, and $P_\text{o}$ can be empirically measured from long-term observations. However, since the coherence time of the random process $\mathcal{S}$ is commonly significantly longer than that of $r_s$, it will take a long time and huge number of observations to efficiently traverse most states in $\mathcal{S}$. During this time, a lot of service outages may have been caused by the resource overload, or even the parameters of the random processes can also have changed. To effectively support the optimization of admission control policy, we need an agiler estimator for $P_{\text{o}}$.

To achieve this, we decompose $P_\text{o}$ into
\begin{equation}\label{eq:overload_risk_decomposition}
	\begin{split}
		P_{\text{o}}&=\sum\limits_{n=0}^{n_{\text{max}}}P\left(\left.\sum\limits_{s\in\mathcal{S}}r_s>r_{\text{max}}\right\vert\vert\mathcal{S}\vert=n\right)P_n\\
		&=\sum\limits_{n=0}^{n_\text{max}}P_n\int\limits_{x=0}^{r_\text{max}}f_n(x)\text{d}x
	\end{split}
\end{equation}

With a $\lambda$-Poisson birth of new slices, an i.i.d. $(1/\mu)$-geometric lifetime of active slices, and an $n_\text{max}$-truncated pool of active slices, $\mathcal{S}$ can be easily identified as a classical $M/M/c/K$ queue where $c=K=n_\text{max}$. We know from the famous Erlang's loss formular that
\begin{equation}\label{eq:erlang_loss}
	P_{n} =\frac{(\lambda/\mu)^n}{n!}\left/\left(\sum\limits_{n=0}^{n_\text{max}}\frac{(\lambda/\mu)^i}{i!}\right)\right. .
\end{equation}

\section{Proposed Approach}\label{sec:approach}

\begin{figure}[!h]
	\centering
	\includegraphics[width=1.13\linewidth]{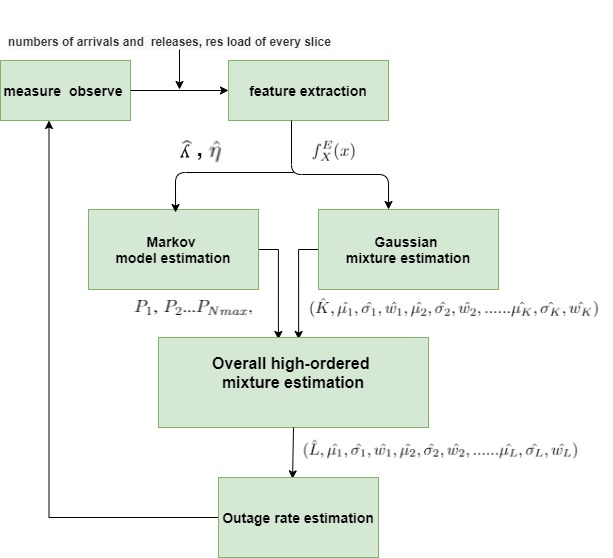}
	\caption{Flow chart of the proposed approach}
	\label{fig: FLOGERT}
\end{figure}

With respect to Eq.~\eqref{eq:overload_risk_decomposition}, we design an estimation approach as shown in Figure \ref{fig: FLOGERT}. By accounting the creation and release of network slices, and continuously monitoring the real-time resource load of every single slice, the MNO is able to extract some basic features from the slice dynamics, including the birth rate $\lambda$, the death rate $\eta$, and an empirical distribution $f_X^\text{E}(x)$ of the resource load in every single slice.

Clearly, $f_X^\text{E}(x)$ can be directly used as an estimate of $f_1(x)$, it is also an option to simulate empirical distributions of multi-slice resource load by randomly sampling $f_X^\text{E}(x)$. However, the coherence among samples may lead to low accuracy of the estimation. Furthermore, raw empirical models in non-parametric form such as histogram or CDF curves are impractical to maintain or analyze, when compared to parameterized distribution models. Therefore, here we propose to take a parametric approach, i.e. to estimate the Gaussian mixture parameters $(K,\mu_1,\sigma_1,\mu_2,\sigma_2,\dots,\mu_K,\sigma_K)$.

It is widely applied to invoke the $K$-means clustering algorithm in the estimation of Gaussian-mixture models. In this approach, all data samples are clustered into $K$ groups, under the assumption that the data values in every individual cluster are Gaussian distributed - the clusters are optimized in such a way that it maximizes the overall likelihood of all sample subordinations. We can then extract a tuple of Gaussian parameters $(\hat\sigma_i,\hat\mu_i)$ from every cluster $i\in\{1,2,\dots,K\}$. However, the $K$-means algorithm requires the value of $K$, which is unknown in the scenario that we study. To overcome this challenge, we first smooth the empirical PDF curve $f_X^\text{E}(x)$ with an moving-average filter, then detect the peaks in the smoothed curve as a rough estimation $\hat{K}$.

Meanwhile, with the estimated birth rate $\hat\lambda$ and estimated death rate $\hat\eta$, which we obtain from the feature extraction, it is easy to calculate $P_1, P_2\dots P_{N_\text{max}}$ by invoking Eq.~\eqref{eq:erlang_loss}.

With the estimated Markov model of active slice set and the Gaussian mixture model of single-slice resource load, we are able to simulate the overall resource load in a Monte-Carlo manner. The obtained empirical distribution is, once again, modeled as a truncated Gaussian-mixture, from which the overload risk is estimated.

\section{Numerical Verification}\label{sec:simulation}

\begin{figure}[!hbtp]
	\centering
	\includegraphics[width=\linewidth]{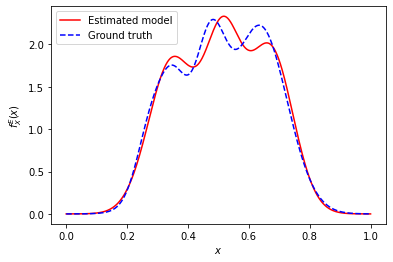}
	\caption{A sample estimate to randomly generated truncated Gaussian mixture distribution.}
	\label{fig:gmm_sample}
\end{figure}

\begin{figure}[!hbtp]
	\centering
	\includegraphics[width=\linewidth]{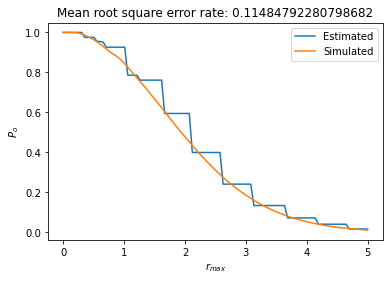}
	\caption{The estimate to overload risk regarding the resource load distribution model in Fig.~\ref{fig:gmm_sample}.}
	\label{fig:outage_risk_sample}
\end{figure}

To verify our proposed approach, we carried out numerical simulations. A one-dimensional resource pool normalized to the size of $5$ is considered to support up to $n_\text{max}=10$ active slices, where every slice's instantaneous resource load is simulated by a truncated mixture model with $5$ randomly generated Gaussian components. More specifically, for every component $i\in\{1,2,3,4,5\}$, the mean $\mu_i\sim\mathcal{U}(0.25,0.75)$, the standard deviation $\sigma_i\sim\mathcal{U}(0,0.1)$, and the weight vector $\mathbf{w}$ is randomly selected and normalized, guaranteeing $\vert\mathbf{w}\vert=1$. The estimator takes $5000$ independent observations of the resource load status to estimate the mixture distribution - and therewith estimate the overload risk $P_\text{o}$. A sample estimate of the truncated Gaussian mixture model is shown in comparison with the ground truth in Fig.~\ref{fig:gmm_sample}, and the corresponding estimate to the resource overload risk is given in Fig.~\ref{fig:outage_risk_sample}.

\begin{figure}[!hbtp]
	\centering
	\includegraphics[width=\linewidth]{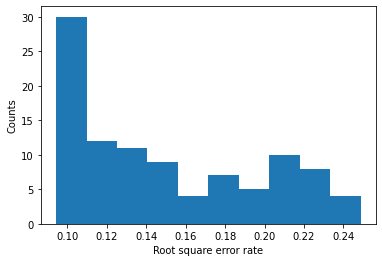}
	\caption{The histogram of estimation error rate over 100 random specifications.}
	\label{fig:error_hist}
\end{figure}

We repeated the test $100$ times, each time with a different random specification of the Gaussian mixture model, each time we measure the mean root square error rate of the estimation to $P_\text{o}$. The error distribution over $100$ tests is illustrated in Fig.~\ref{fig:error_hist}, showing a satisfactory accuracy.

\begin{figure}[!hbtp]
	\centering
	\includegraphics[width=\linewidth]{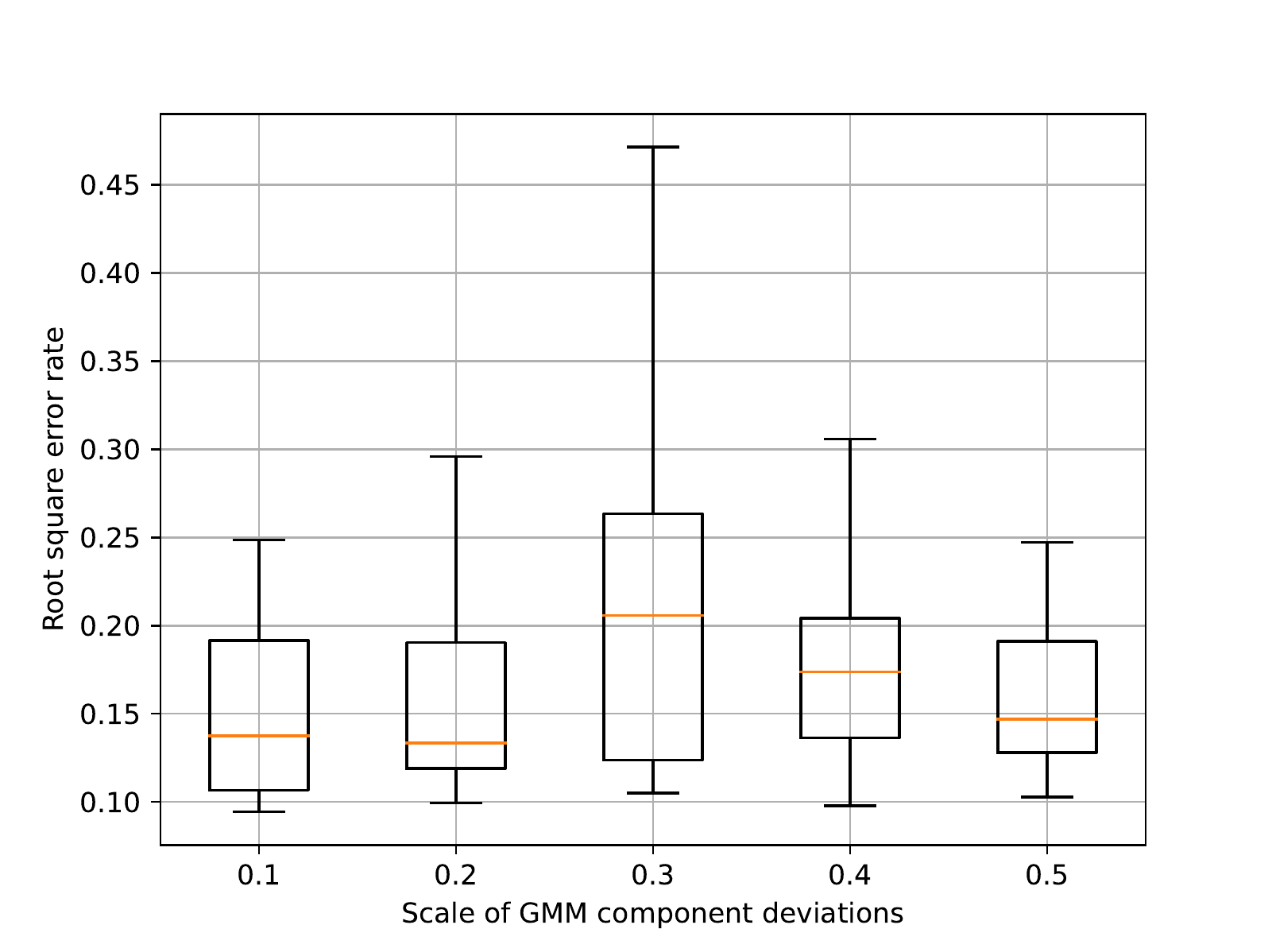}
	\caption{The estimation error rate with different deviations of Gaussian components in the GMM model.}
	\label{fig:GMM_deviations_sensitivity}
\end{figure}

Furthermore, for a deeper insight we tested the sensitivity of our approach to different aspects of the slice model, including the number of Gaussian components of the GMM model, and the scale of Gaussian parameters $(\mu_i,\sigma_i)$, respectively. More specifically, we repeated the aforementioned Monte-Carlo tests with varied specifications, and plot the distributions of resulted estimation errors, illustrated in Figs.~\ref{fig:GMM_components_sensitivity}--\ref{fig:GMM_deviations_sensitivity}, respectively.
Generally, we observe the error to increase along with the number GMM components, which may root in the increased errors in Gaussian component detection, and the accumulated error in estimating each additional Gaussian component. With respect to the scale of Gaussian parameters, the error appears to increase first, and then vanishes out after hitting a peak level, which ensures a wide applicability of our proposed method in various load scenarios.

\begin{figure}[!hbtp]
	\centering
	\includegraphics[width=\linewidth]{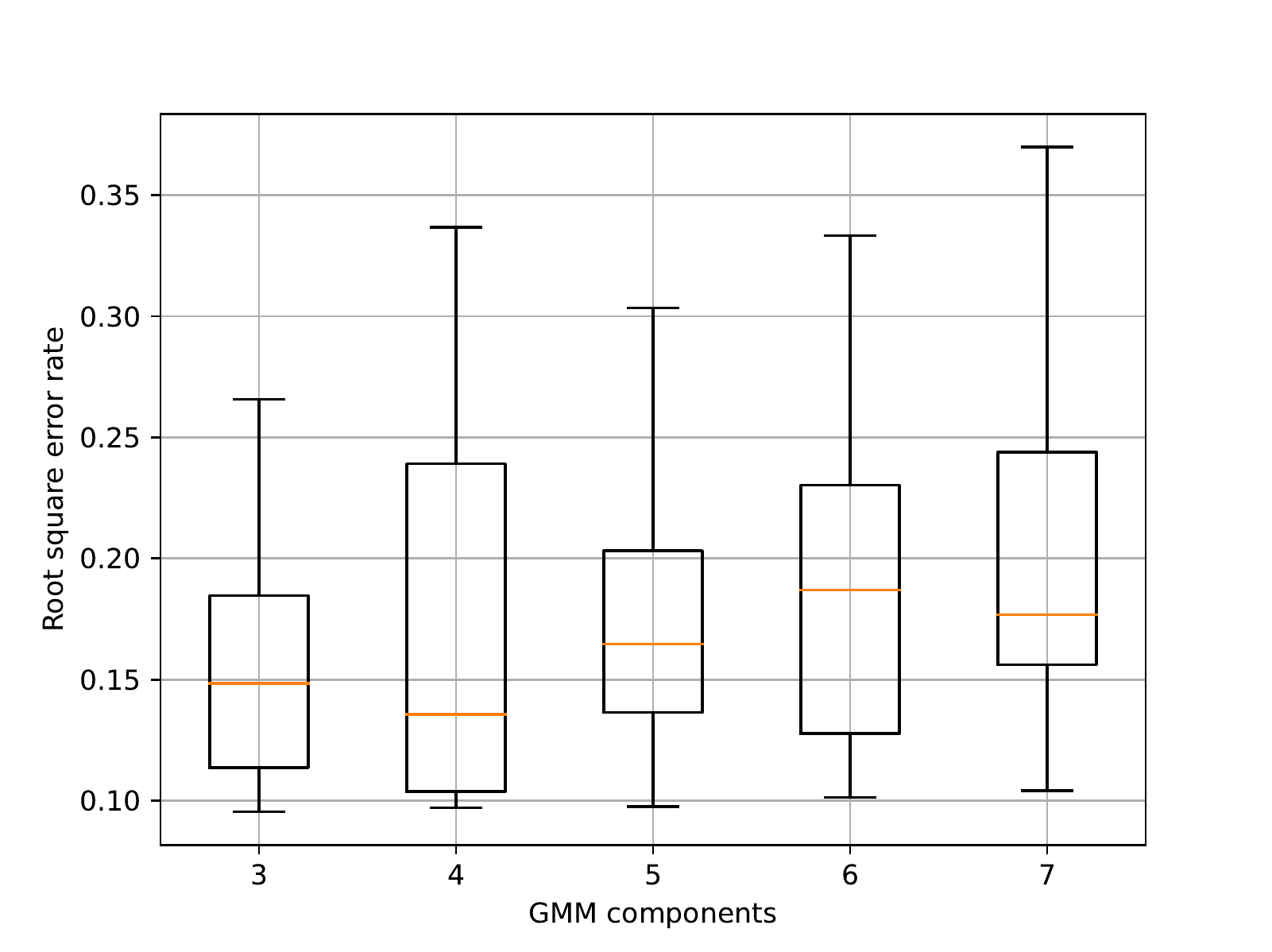}
	\caption{The estimation error rate with different numbers of Gaussian components in the GMM model.}
	\label{fig:The estimation error rates over 100 random specifications.}
\end{figure}

\begin{figure}[!hbtp]
	\centering
	\includegraphics[width=\linewidth]{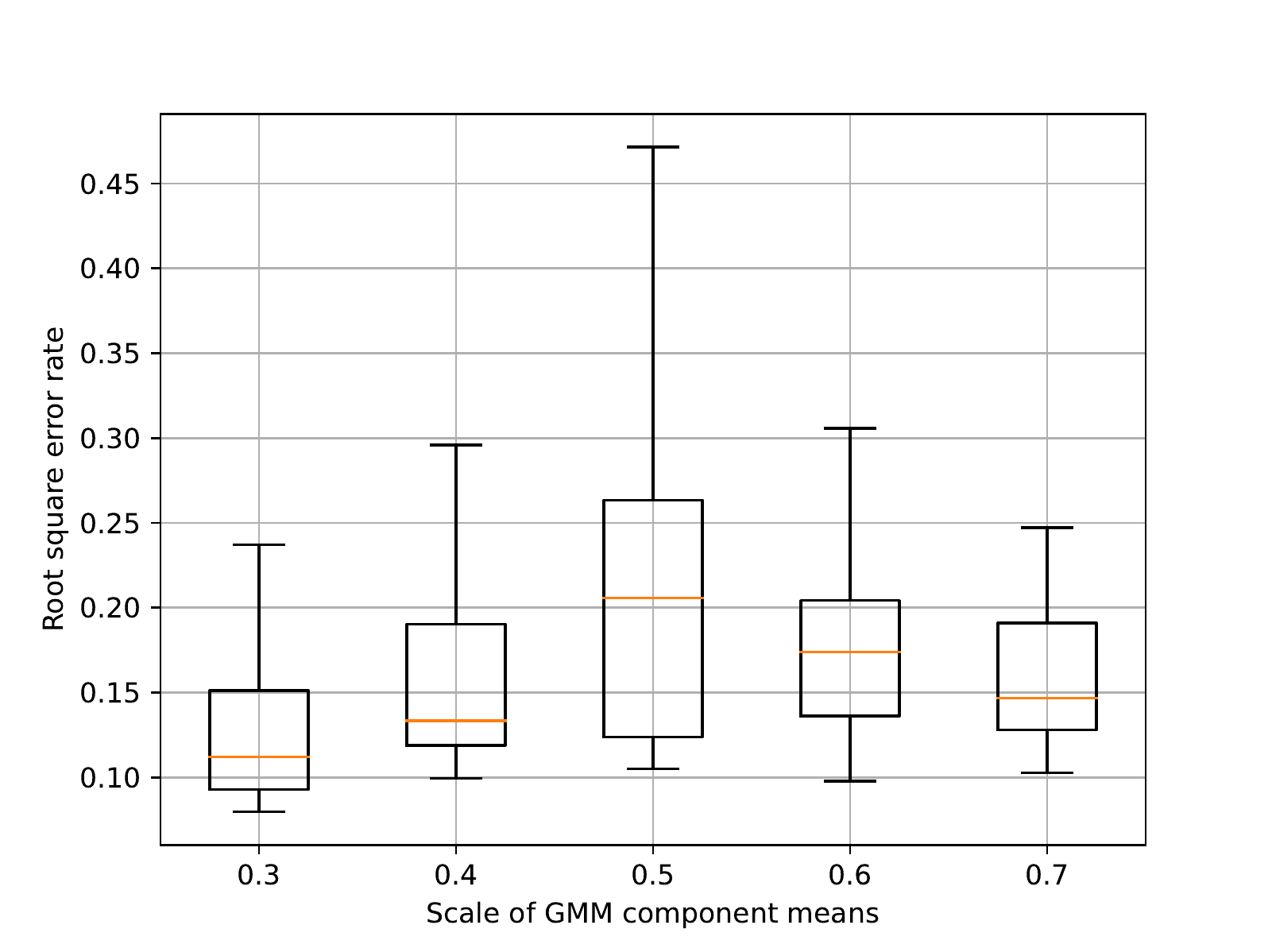}
	\caption{The estimation error rate with different means of Gaussian components in the GMM model.}
	\label{fig:GMM_means_sensitivity}
\end{figure}

\section{Conclusion}\label{sec:conclusion}

In this paper we have presented a estimator of resource overload risk in 5G multi-tenancy networks with homogeneous slices, which have i.i.d. truncated Gaussian-mixture random instantaneous resource load. The proposed method uses $K$-means clustering algorithm on observations of single-slice resource load, to extract the Gaussian components from an unknown mixture distribution. Then we assemble the estimated mixture model with a $M/M/c/K$ queuing model, to construct the high-order mixture model for overall resource load based on Monte-Carlo approach. The proposed method is verified by numerical tests as effective and accurate. For future work, we are looking forward to integrating this overload risk estimator into slice admission control solutions, in order to supervise a better admission policy optimization that is sensitive to the overload risk in elastic slicing.

\bibliographystyle{chicago}
\bibliography{bibliography}

\end{document}